\documentclass[10pt,prd,psfig,dcolumn,epsfig,superscriptaddress]{revtex4}
\usepackage{graphicx}
\usepackage{multirow}
\usepackage{amsmath}
\usepackage{slashed}
\begin{document}
\title{Hidden-charm Pentaquark Production at $e^+e^-$ Colliders}
\author{Shi-Yuan Li}
\affiliation{School of Physics, Shandong University, Jinan 250100, P. R. China}
\author{Yan-Rui Liu}
\affiliation{School of Physics, Shandong University, Jinan 250100, P. R. China}
\author{Yu-Nan Liu}
\affiliation{School of Physics, Shandong University, Jinan 250100, P. R. China}
\author{Zong-Guo Si}
\affiliation{School of Physics, Shandong University, Jinan 250100, P. R. China}
\affiliation{Institute of Theoretical Physics, Chinese Academy of Science, Beijing 100190, P. R. China}
\author{Xiao-Feng Zhang}
\affiliation{School of Physics, Shandong University, Jinan 250100, P. R. China}

\date{\today}

\begin{abstract}
We study one possible production mechanism for hidden charm pentaquark in $e^+e^-$ collision, where  it is produced  via a color-octet $c\bar c$ pair fragmentation.
  %
The  pentaquark production at B factory energy  is dominated by  $e^+e^- \to c \bar c g \to P_c+X$.
At $Z^0$ pole, for the  pentaquark production, there are several  partonic processes  playing significant role.   Our results show that it is possible to
search for the direct pentaquark production signal at $e^+e^- $  colliders,  which    is important to understand the properties of pentaquark.
\end{abstract}

\maketitle

To study the properties and production mechanisms of the multiquark state is important to understand the quark model and the strong interactions deeply.  The newest evidence for the existance of the pentaquark is from LHCb experiments. Recently, the LHCb Collaboration announced the observation of two charged hidden-charm resonances $P_c^+(4380)$ and $P_c^+(4450)$ produced in the process $\Lambda_b^0\to K^-J/\psi p$ \cite{Aaij:2015tga,Aaij:2016phn,Aaij:2016ymb}. The corresponding decay channel   $P_c\to J/\psi p$ indicates that their minimal quark content is $uudc\bar{c}$. The most important work at present is to confirm whether the new resonances are pentaquarks or not. Up to now, lots of theoretical investigations have been finished. Several possibilities for these new resonances to be baryon-meson molecules, two- or three-cluster compact states, etc have been discussed \cite{Chen:2016qju,Chen:2016heh}. In fact, studies on hidden-charm pentaquarks have been started before the observation of the $P_c$ states. In Ref. \cite{Wu:2010jy}, the interaction between a charmed baryon and an anticharmed meson was studied and it was found that hadronic molecules with the mass above 4 GeV are possible. While in \cite{Wu:2010vk,Yang:2011wz,Wang:2011rga,Yuan:2012wz,Wu:2012md,Garcia-Recio:2013gaa,Xiao:2013yca,Uchino:2015uha,Huang:2013mua,Wang:2015xwa,Garzon:2015zva,Li:2014gra}, the possibility for this kind of bound state to be the hidden-heavy pentaquark was studied.

In the mean time, the investigations on the pentaquark production in various collisions have also been performed. In Ref. \cite{Huang:2013mua}, the discovery potential of hidden-charm pentaquarks in the photon-induced production was discussed. In Ref. \cite{Wang:2015jsa}, the $J/\psi$ photoproduction of the two $P_c$ states off the proton was proposed to understand their nature. Refs. \cite{Kubarovsky:2015aaa,Karliner:2015voa,Huang:2016tcr,Blin:2016dlf,Kubarovsky:2016whd} also present the pentaquark production in $\gamma N$ collision processes. In Refs. \cite{Garzon:2015zva,Lu:2015fva,Kim:2016cxr,Ouyang:2015rre}, the hidden-charm pentaquark effects in the $\pi^-p$ reaction were considered. Discussions for the production in bottom baryon decays \cite{Cheng:2015cca,Hsiao:2015nna,Ali:2016dkf} and in heavy-ion or $pA$ collisions \cite{Wang:2016vxa,Schmidt:2016cmd} can also be found.

However, most of the discussions on production in the present literature are at hadron level and the understanding for the pentaquark structures needs more studies. Obviously, the $P_c^+(4380)$ is not a bound state of $J/\psi$ and nucleon. A compact $uudc\bar{c}$ pentaquark state with colored $c\bar{c}$ may be formed through gluon-exchange interactions. The spectrum and qualitative decay properties of the compact pentaquarks \cite{Wu:2017weo,Takeuchi:2016ejt} indicate that such a configuration is not contradicted with the observed $P_c$ states. In principle, various configurations of a five-quark system require different production mechanisms. For the hadron-hadron molecules, the produced quarks fragment firstly into various hadrons and then the residual strong interactions between these hadrons lead to possible hadronic molecules. One may study the production at hadron level \cite{Jin:2016vjn}. For the compact $uudc\bar{c}$ pentaquark state with colored $c\bar{c}$, a feasible approach is the framework proposed in Ref. \cite{Ma:2003zk}. Since the gluon is easily converted to a colored charm-anticharm pair, the production rate of the considered compact pentaquark might be significant. We here discuss the production of such a type of pentaquark in the multiproduction process which can help to better understand their structure as well as the strong interaction mechanism at the hadronization scale \cite{Jin:2016cpv}. The information of the cross section, rapidity and transverse momentum distributions, {\it etc.}, of the relevant particles on a specific collider can help the experimentalists to set the proper triggers and cutoffs for the measurements \cite{Jin:2016vjn, Jin:2014nva}. Among the high energy collisions, the $e^+$ $e^-$ annihilation process is of special advantage for its clean background and one can gain more clear picture on the colour and other structure evolution via the study of the production.

For this kind of hidden-charm pentaquark states, we can rely on the perturbative QCD (PQCD) to calculate the charm quark pair production. On the other hand,  how to embed the PQCD result into the production amplitude of the pentaquark, depends on the structure of the state and the framework for the approximation. Here we ignore the consideration of it as hadron bound state, which has been studied in the general case \cite{Jin:2016vjn}. A benchmark framework could be the heavy quark effective theory, which provides the feasible factorization formulation to connect the PQCD process with the parameterization of the nonperturbative QCD process, i.e., the PQCD produced $c\bar{c}$ transiting to the pentaquark, in the rest frame of the bound state. For concrete, we employ the nonrelativistic QCD (NRQCD) factorization framework. The NRQCD factorization approach has been used to discuss the production of $\Xi_{cc}$ in Ref. \cite{Ma:2003zk} and that of $T_{cc}$ in Ref. \cite{HyodoLOSY2013} at various $e^+e^-$ colliders. The aim of this paper is to study the hidden-charm pentaquark production in
the process
\begin{eqnarray}
e^-(p_1)+e^+(p_2)\to \gamma^*/Z^0 \to P_c(k)+X,
\end{eqnarray}
where $p_1$, $p_2$, and $k$ denote the momenta of the related particles. The unobserved part $X$ can  always be divided into a perturbative part $X_P$ and a nonperturbative part $X_N$, $X=X_N+X_P$.
The corresponding invariant amplitude  can be written as
\begin{eqnarray} \label{amplitude}
\mathcal{M}=\int\frac{d^4k_1}{(2\pi)^4}A_{ij}(k_1,k_2,p)\int d^4x_1e^{-ik_1\cdot x_1}\langle P_c(k)+X_N|\bar{Q}_i(x_1)Q_j(0)|0\rangle,
\end{eqnarray}
where both $i$ and $j$ take the Dirac and color indices. $Q(x)$ is the Dirac field for charm quark and $p$ denotes the total momentum of the partons appearing in the perturbative part. If $\langle P_c(k)+X_N|$ is replaced by the state of a free charm-anticharm quark pair with momenta $k_1$ and $k_2$, ${\cal M}$ is the amplitude for the process $e^+ e^- \to \gamma^*/Z^0 \to c(k_1)+\bar{c}(k_2)+g(k_g)$ with $p=k_g$, so the lowest order contribution of this process is the production of the color-octet charm quark-antiquark pair and then this pair fragments into the pentaquark $P_c$.

\begin{figure}[htb]
\centering
\scalebox{0.45}{\includegraphics{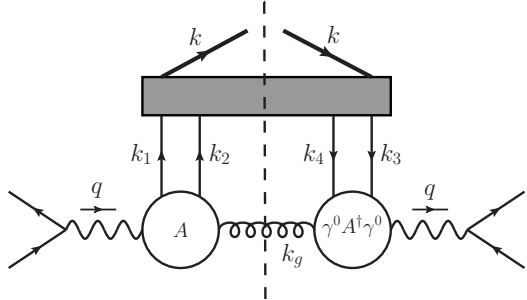}}
\caption{Graphic representation for the contribution in Eq. (\ref{MatrixElment})}\label{Lowest}
\end{figure}

With the above amplitude ${\cal M}$, the cross section for the process $e^+e^-\to \gamma^*/Z^0\to c\bar{c}g\to P_c+X$ can be written as
\begin{eqnarray}\label{eq3}
d\sigma&=&\frac{1}{2s} {\frac{1}{4}}\frac{d^3k}{(2\pi)^3}\frac{d^3k_g}{(2\pi)^3 2E_g}(2\pi)^4 \delta^4(p_1+p_2-k-k_g)\nonumber\\
&&\times\int\frac{d^4k_1}{(2\pi^4)}\frac{d^4k_3}{(2\pi^4)} A_{ij}(k_1,k_g)(\gamma^0A^\dag(k_3,k_g)\gamma^0)_{kl}\nonumber\\
&&\times\int d^4x_1d^4x_3e^{-ik_1\cdot x_1+ik_3\cdot x_3}\langle0|\bar{Q}_k(0)Q_l(x_3)|P_c+X_N\rangle\langle P_c+X_N|{Q}_i(x_1)\bar{Q}_j(0)|0\rangle.
\end{eqnarray}
Here we take nonrelativistic normalization for $P_c$. The spin average of initial leptons, spin summation of final $P_c$, and the polarization and color summation of gluon are implied. By using translational covariance one can eliminate the summation over $X_N$. Defining the creation operator $a^\dag({\bf k})$ for $P_c$ with the three momentum ${\bf k}$, we obtain
\begin{eqnarray}\label{MatrixElment}
d\sigma&=&\frac{1}{2s}\frac{1}{4}\frac{d^3k}{(2\pi)^3} \frac{d^3k_g}{(2\pi)^3 2E_g}\int\frac{d^4k_1}{(2\pi^4)}\frac{d^4k_3}{(2\pi^4)} A_{ij}(k_1,k_2,p_3,p_4)(\gamma^0A^\dag(k_3,k_4,p_3,p_4)\gamma^0)_{kl}\nonumber\\
&&\times\int d^4x_1d^4x_2d^4x_3 e^{-ik_1\cdot x_1-ik_2\cdot x_2+ik_3\cdot x_3}\langle0|\bar{Q}_k(0)Q_l(x_3)a^\dag({\bf k})a({\bf k}){Q}_i(x_1)\bar{Q}_j(x_2)|0\rangle.
\end{eqnarray}
This is shown by Fig. \ref{Lowest}, where the black box represents the Fourier transformed matrix element of the second line in Eq.(\ref{MatrixElment}).

Since heavy quarks move with a small velocity $v_Q$ inside the pentaquark
in its rest frame, the Fourier transformed matrix element can be expanded in $v_Q$ with fields
of NRQCD. The relation between NRQCD fields and Dirac field $Q(x)$
in  $P_c$'s rest frame is
\begin{equation} \label{eq5}
Q(x)= e^{-im_Q t} \left\{\begin{array}{ll} \psi(x) \\ \ 0 \end{array}\right\} +
e^{im_Q t} \left\{\begin{array}{ll} \ 0 \\ \chi(x)\end{array}\right\} +
{\cal O}(v_Q),
\end{equation}
where $\psi(x)$($\chi(x)$) denotes the Pauli spinor field that annihilates (creates) a heavy (anti-)quark.
We will work at the leading order of $v_Q$. In order to express our results for the Fourier transformed matrix element in a covariant way, we employ the four-velocity of the pentaquark with $v_{\mu} = k_{\mu}/M_{P_c}$. The Fourier transformed matrix element is related to that in the rest frame \cite{Ma:2003zk}:
\begin{eqnarray}\label{eq6}
&& v^0 \int d^4x_1 d^4x_2 d^4x_3 e^{-ik_1\cdot x_1-ik_2\cdot x_2 +ix_3\cdot k_3}
   \langle 0\vert  \bar{Q}_k(0) Q_l(x_3)  a^\dagger ({\bf k}) a({\bf k}) Q_i(x_1) \bar Q_j(x_2) \vert 0\rangle \nonumber\\
&& = \int d^4x_1 d^4x_2 d^4x_3 e^{-ik_1\cdot x_1-ik_2\cdot x_2 +ix_3\cdot k_3}
   \langle 0\vert  \bar{Q}_k(0) Q_l(x_3)  a^\dagger ({\bf k}=0)a({\bf k}=0)  Q_i(x_1) \bar Q_j(x_2) \vert 0\rangle.
\end{eqnarray}

Using Eq. (\ref{eq5}), one can expand the matrix element in Eq. (\ref{eq6}) with $\psi^\dagger (x)$ and $\chi(x)$.
The space-time of the matrix element with NRQCD fields is controlled by the scale $m_Q v_Q$. Hence at leading order of $v_Q$ one can neglect the space-time dependence in $\psi^\dagger (x)$ and $\chi(x)$. With this approximation,
the matrix element in Eq.(\ref{eq6}) is
\begin{equation}
\langle 0\vert  \psi^{a_3\dagger}_{\lambda_3}(0) \chi^{a_4}_{\lambda_4}(0)  a^\dagger
       a \chi^{a_1}_{\lambda_1}(0) \psi^{a_2\dagger}_{\lambda_2}(0) \vert 0\rangle,
\end{equation}
where we suppressed the notation ${\bf k}=0$ in $a$ and $a^\dag$ and it is always implied
that NRQCD matrix elements are defined in the rest frame of $P_c$. The superscripts $a_{i}(i=1,2,3,4)$ are for the color of quark fields, while the subscripts $\lambda_i (i=1,2,3,4)$ for the quark spin indices. Within NRQCD, we obtain
\begin{eqnarray}
\langle 0\vert  \psi^{a_3\dagger}_{\lambda_3}(0) \chi^{a_4}_{\lambda_4}(0)  a^\dagger
       a \chi^{a_1}_{\lambda_1}(0) \psi^{a_2\dagger}_{\lambda_2}(0) \vert 0\rangle
  = \delta_{\lambda_4\lambda_3} \delta_{\lambda_2\lambda_1} T^c_{a_3a_4} T^c_{a_1a_2} h_1
   +\sigma^n_{\lambda_4\lambda_3}\sigma^n_{\lambda_2\lambda_1} T^c_{a_1a_2}T^c_{a_3a_4} h_3 +  cst,
\end{eqnarray}
where $\sigma^n (n=1,2,3)$ are Pauli matrices, $2T^c(c=1,\cdots,8)$ are Gell-Mann matrices, and the color-singlet term (cst) is irrelevant here. The parameters $h_1$ and $h_3$ are defined as:
\begin{eqnarray}
h_1 &=& \frac{1}{8}
   \langle 0\vert  (\psi^{\dagger} T^c \chi)\ a^\dagger a \ (\chi^{\dagger} T^c \psi) \vert 0\rangle,
\nonumber\\
h_3 &=& \frac{1}{16}
   \langle 0\vert (\psi^{\dagger} T^c \sigma^n \chi) \
       a^\dagger a \ (\chi^{\dagger}  T^c \sigma^n \psi) \vert 0\rangle.
\end{eqnarray}
$h_1$($h_3$) represents the probability for a color-octet $Q\bar{Q}$ pair in a $^1S_0$($^3S_1$) state to transform into the pentaquark. With these results, the Fourier transformed matrix element in Eq.(\ref{eq6}) can be expressed as
\begin{eqnarray}\label{eqv0}
&&v^0\int d^4x_1d^4x_2d^4x_3 e^{-ik_1\cdot x_1-ik_2\cdot x_2+ik_3\cdot x_3}\langle0|
\bar{Q}_k(0)Q_l(x_3)a^\dag({\bf k})a({\bf k})\bar{Q}_i(x_1)Q_j(x_2)|0\rangle \nonumber\\
&&~~~=(2\pi)^4\delta^4(k_1-m_Qv)(2\pi)^4\delta^4(k_2-m_Qv)(2\pi)^4\delta^4(k_3-m_Qv)\nonumber\\
&&~~~~~~\times\bigg[-(T_{a_3a_4}^aT_{a_1a_2}^a)\cdot(P_-\gamma_5P_+)_{ji}(P_+\gamma_5 P_-)_{lk} \cdot h_1\nonumber\\
&&~~~~~~~~~~~+(T_{a_3a_4}^aT_{a_1a_2}^a) \cdot (P_-\gamma^\mu P_+)_{ji}(P_+\gamma^\nu P_-)_{lk}(v_\mu v_\nu-g_{\mu\nu} ) \cdot h_3 \bigg],
\end{eqnarray}
where $P_\pm=\frac{1\pm\gamma \cdot v}{2}$.

With the above formula, we obtain the differential cross section as follows
\begin{align}
  d\sigma &=\frac{1}{8s} \int \frac{d^3 k}{(2\pi)^3 v_0} \int \frac{d^3 k_g}{(2\pi)^32E_g}A_{ij}(k,k_g)\Big[\gamma^0A^+(k,k_g)\gamma^0\Big]_{kl}
  (2\pi)^4\delta^4(p_1+p_2-k-k_g)\nonumber\\
&~~~~~\bigg\{-(P_-\gamma_5P_+)_{ji}(P_+\gamma_5P_-)_{lk}\cdot T_{a_3a_4}^aT_{a_1a_2}^ah_{1}\nonumber\\
&~~~~~+\Big[(P_-\slashed vP_+)_{ji}(P_+\slashed vP_-)_{lk}-(P_-\gamma_{\sigma}P_+)_{ji}(P_+\gamma_{\sigma}P_-)_{lk}\Big]\cdot
T_{a_3a_4}^aT_{a_1a_2}^ah_{3} \bigg\}.
\end{align}
\begin{figure}[!h]
\centering
\scalebox{0.4}{\includegraphics{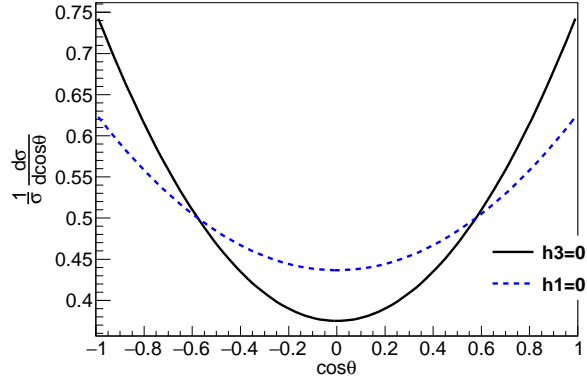}}
\caption{Angular distribution $\frac{1}{\sigma} \frac{d\sigma}{d\cos\theta}$  of $e^-(p_1)+e^+(p_2)\to \gamma^*/Z^0 \to P_c(k)+g$ process at $\sqrt{s}=10.6$ GeV. The dash line is for $h_1=0$, and the solid one is for $h_3=0$. }
\label{LOcosG10}
\end{figure}
The angular distribution $\frac {1}{\sigma}\frac{d\sigma}{d\cos\theta}$ of the pentaquark $P_c$ is shown in Fig. \ref{LOcosG10}, where $\theta$ is the angle between the moving direction of the $e^-$ beam and that of $P_c$.

After integrating over the phase space, we obtain the total cross section for the pentaquark production at the lowest order
\begin{align}
\sigma=\frac{128\pi\alpha_s\sigma_0}{(4m_c^2+M_P^2-2s)^2}(1-\frac{M_P^2}{s})\left[\mathcal{A}~h_1+\mathcal{B}~h_3\right],
\end{align}
where $M_P$ is the pentaquark mass, $\sigma_0=4\pi\alpha^2/3s$, $\alpha$ ($\alpha_s$) is the electromagnetic (strong) coupling constant, and
\begin{align}
   \mathcal{A}=\frac{4(s-M_P^2)^2}{M_P}\left\{ \frac{Q_c^2}{s}+\frac{2V_eV_cQ_c(s-M_Z^2)}{(s-M_Z^2)^2+\Gamma_Z^2M_Z^2}
   +\frac{s(A_e^2+V_e^2)V_c^2}{(s-M_Z^2)^2+\Gamma_Z^2M_Z^2} \right\},
\end{align}
\begin{align}
   \mathcal{B}=\frac{3A_c^2(A_e^2+V_e^2)}{M_P[(s-M_Z^2)^2+\Gamma_Z^2M_Z^2]}
   \left\{ \Big[M_P^3-3sM_P+2m_c(s+M_P^2)\Big]^2 +\frac{2}{3}(s-M_P^2)^2\Big[2s-(2m_c+M_P)^2\Big] \right\}.
\end{align}
Here, $M_Z$ ($\Gamma_Z$) is the mass (width) of the $Z^0$ boson, $Q_c$ is the electric charge of the charm quark, $A_e$, $V_e$, $A_c$, and $V_c$ for electron and $c$ quark are

\[
  \left\{
    \begin{aligned}
        &V_e=\frac{1}{sin2\theta_w}(-\frac{1}{2}+2sin^2\theta_w), \\
        &A_e=-\frac{1}{2sin2\theta_w}, \\
        &V_c=\frac{1}{sin2\theta_w}(\frac{1}{2}-\frac{4}{3}sin^2\theta_w),\\
        &A_c=\frac{1}{2sin2\theta_w},
    \end{aligned}
  \right.
\]
where $sin^2\theta_w=0.23$ is adopted for weak-mixing angle.

Obviously, only the weak interactions contribute to the coefficient of $h_3$ due to Furry theorem. If $h_1$ and $h_3$ are at the same magnitude order, at B factory energy, the contribution to the pentaquark production from $h_1$ is dominant, while at $Z^0$ pole, that from $h_3$ is dominant. To show this clearly, we choose  $m_c=1.6$ GeV, $M_Z=91.2$ GeV, $\Gamma_Z=2.5$ GeV, and $M_P=4.38$ GeV, integrate over $\cos\theta$, and then obtain
\begin{align}
   \sigma&=\Big\{42.43\Big[ \frac{h_1}{(\mathrm{GeV})^3}\Big] + 0.0033\Big[\frac{h_3}{(\mathrm{GeV})^3}\Big] \Big\}\  \mathrm{pb} \ \ \ \  \text{for B factory energy,}\nonumber\\
   \sigma&=\Big\{0.39\Big[ \frac{h_1}{(\mathrm{GeV})^3}\Big]+ 2.64 \Big[\frac{h_3}{(\mathrm{GeV})^3}\Big]\Big\}\  \mathrm{pb} \ \ \ \ \ \ \ \ \  \text{for}~Z^0~\text{pole}
\end{align}
where $\alpha(M_Z)=1/128$ and $\alpha_s(M_Z)=0.118$ are adopted.

Till now, $h_1$ and $h_3$ are still unkown exactly. One can  attempt to relate the nonperturbative factors $h_1$ and $h_3$ to the wave function of the $c\bar{c}$ in the pentaquark, e.g.
\begin{equation}\label{hdefine}
h_3 =c_3 \frac{1}{4\pi} \vert R_{c\bar{c}}(0)\vert^2,
\end{equation}
where $R_{c\bar{c}}(\xi)$ represents the radial wave function of the harmonic oscillator composed of $c\bar c$ system and $\xi$ is the Jacobi coordinate for $c\bar{c}$ \cite{HyodoLOSY2013}. In the heavy quark limit, $h_1$ and $h_3$ are identical up to the corrections of order $v_Q^2$, i.e., $h_3=h_1[1+{\cal O}(v_Q^2)]$ \cite{Bodwin:1994jh}. The coefficient $c_3$ is due to the fact that $h_3$ is defined by a color-octet matrix element. In the non-perturbatuive process, the transition from a colour-octet state to the final color-singlet pentaquark state 
introduces extra suppressions of certain power of the relative velocity between the heavy pair in their rest frame, $v_Q$, which we expand the amplitude around its zero value as done above. So the matrix element is suppressed by the small factor proportional to powers of $v_Q$ \cite{Bodwin:1994jh}. In the following  numerical  calculations we simply take $c_3$ as $10^{-1}$ for the charm sector.

The Schrodinger wave function $R_{c\bar{c}}(\xi)$ can be calculated in potential models. The Hamiltonian can be written as \cite{HyodoLOSY2013}
\begin{eqnarray}
\hat{H}_{\xi}=-\frac{1}{2\mu}\nabla_{\xi}^2+\frac{k_{\xi}}{2}{\xi}^2,
\end{eqnarray}
where $\mu=\frac{1}{2}m_c$ is the reduced mass of $c\bar c$.
The wave function of the ground state at origin can be written as
\begin{eqnarray}
R_{c\bar c}(0)=(\frac{7}{16}\mu k_x)^{3/8}\Big[\frac{4}{\sqrt{\pi}}\Big]^{1/2},
\end{eqnarray}
where $k_x=0.33$ GeV$^3$ \cite{HyodoLOSY2013}. From the above equations, we take $h_1=h_3=0.0036$ GeV$^3$ and obtain $\sigma=0.153$ pb for the B factory energy and $\sigma=0.011$ pb for the $Z^0$ pole, respectively.


\begin{figure}[htb]
  \centering
{
    \label{fig:subfig:3} 
    \includegraphics[width=0.35\textwidth]{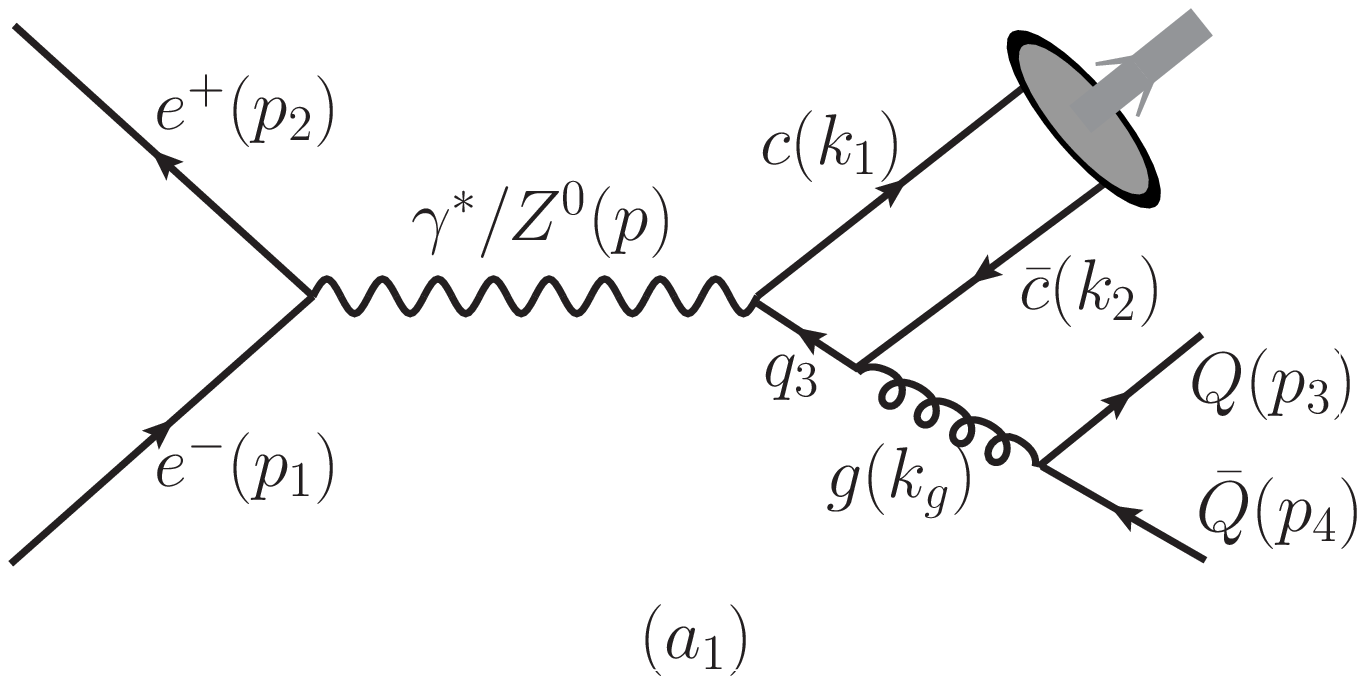}}
  \hspace{1in}
{
    \label{fig:subfig:4} 
    \includegraphics[width=0.35\textwidth]{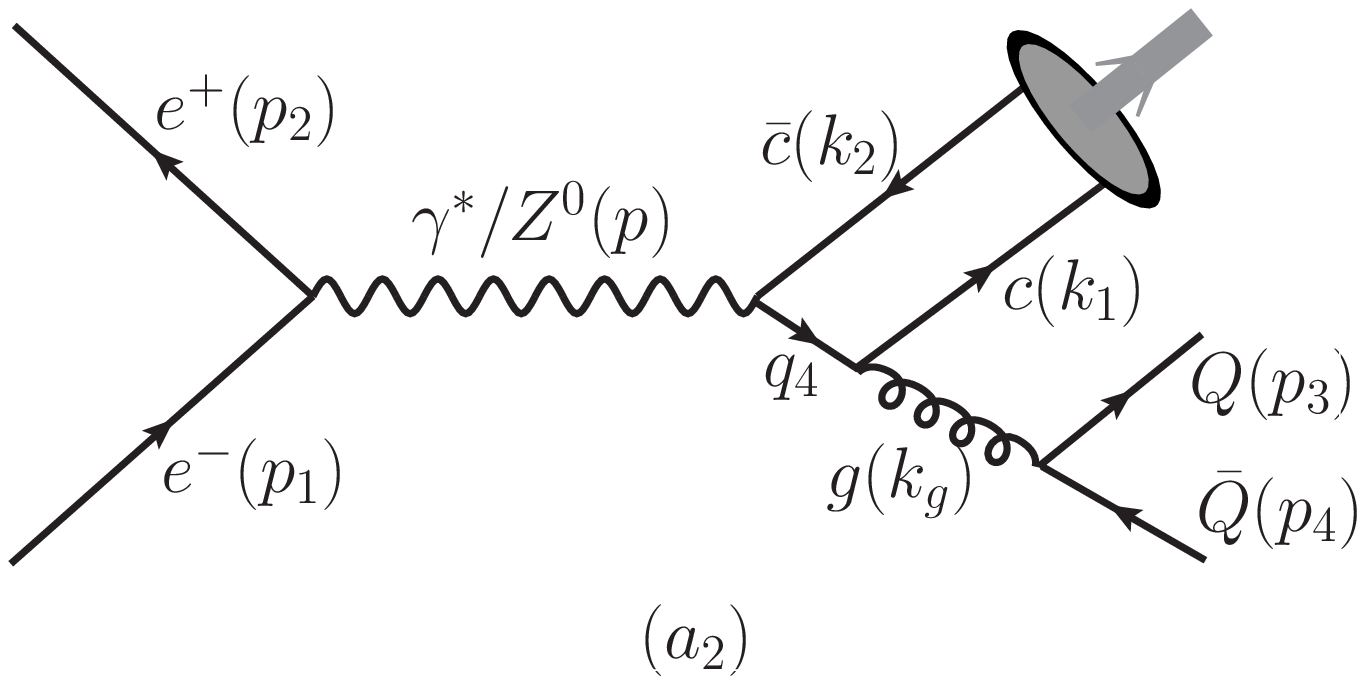}}
{
    \label{fig:subfig:5} 
    \includegraphics[width=0.35\textwidth]{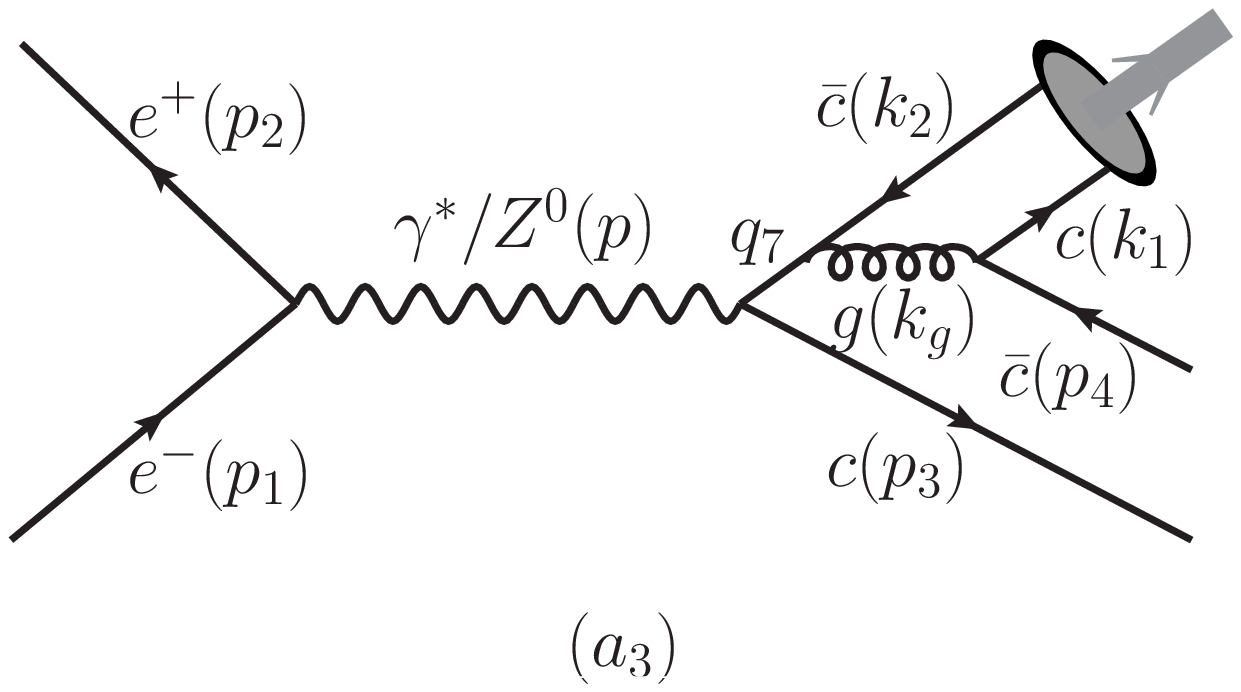}}
  \hspace{1in}
{
    \label{fig:subfig:6} 
    \includegraphics[width=0.35\textwidth]{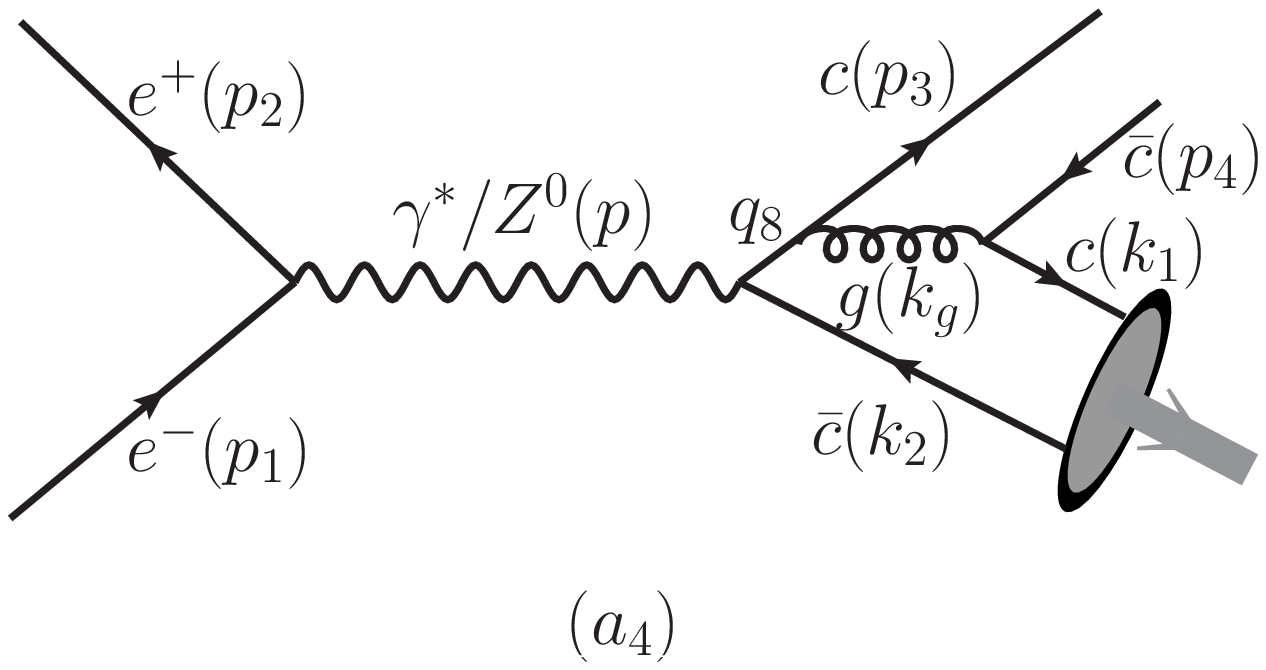}}
  \hspace{1in}
{
    \label{fig:subfig:7} 
    \includegraphics[width=0.35\textwidth]{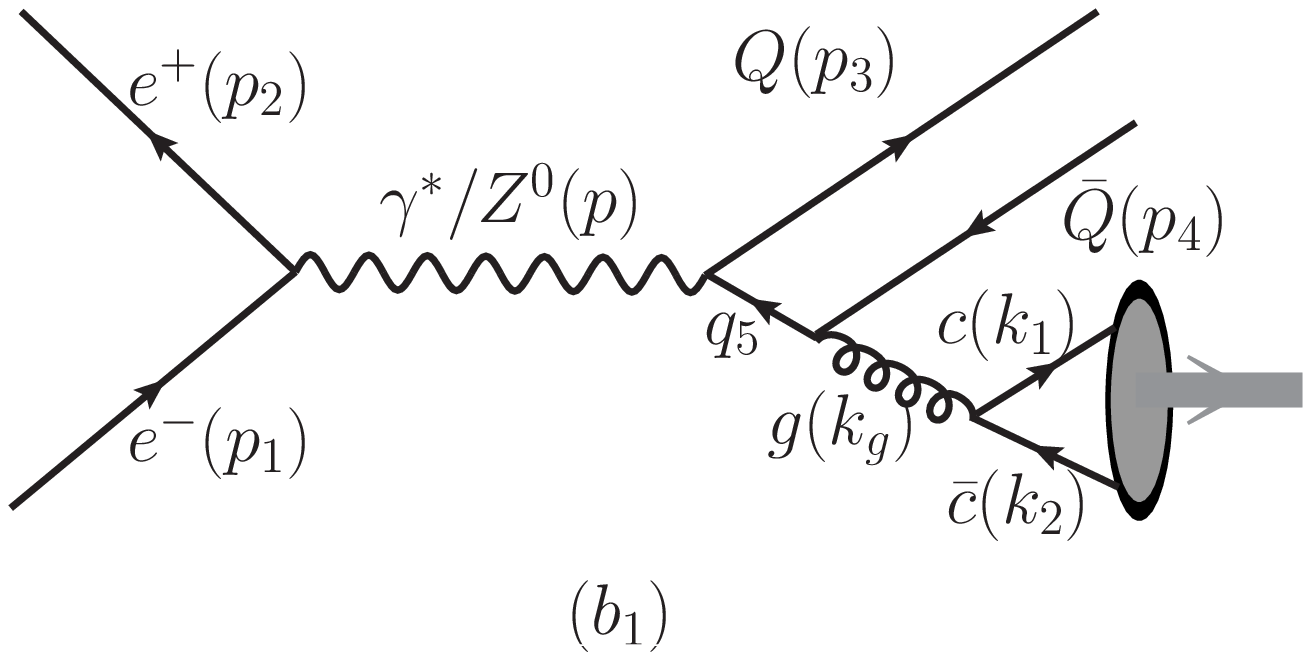}}
  \hspace{1in}
{
    \label{fig:subfig:8} 
    \includegraphics[width=0.35\textwidth]{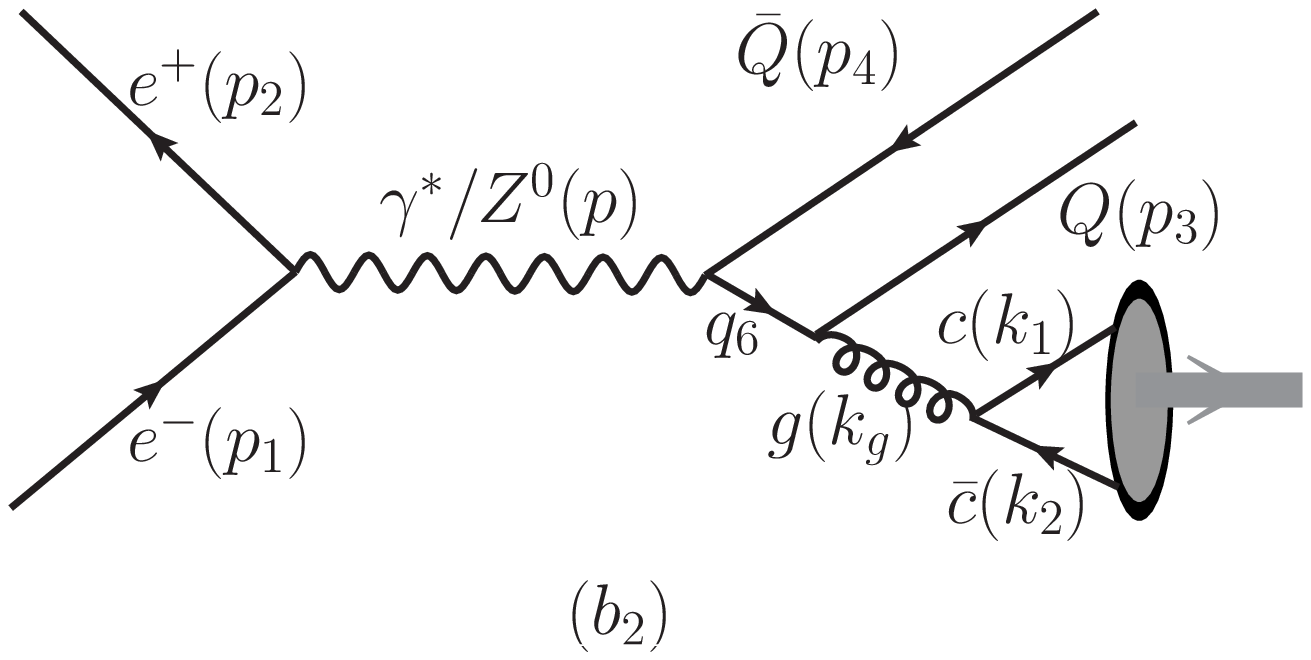}}
  \hspace{1in}
{
    \label{fig:subfig:9} 
    \includegraphics[width=0.35\textwidth]{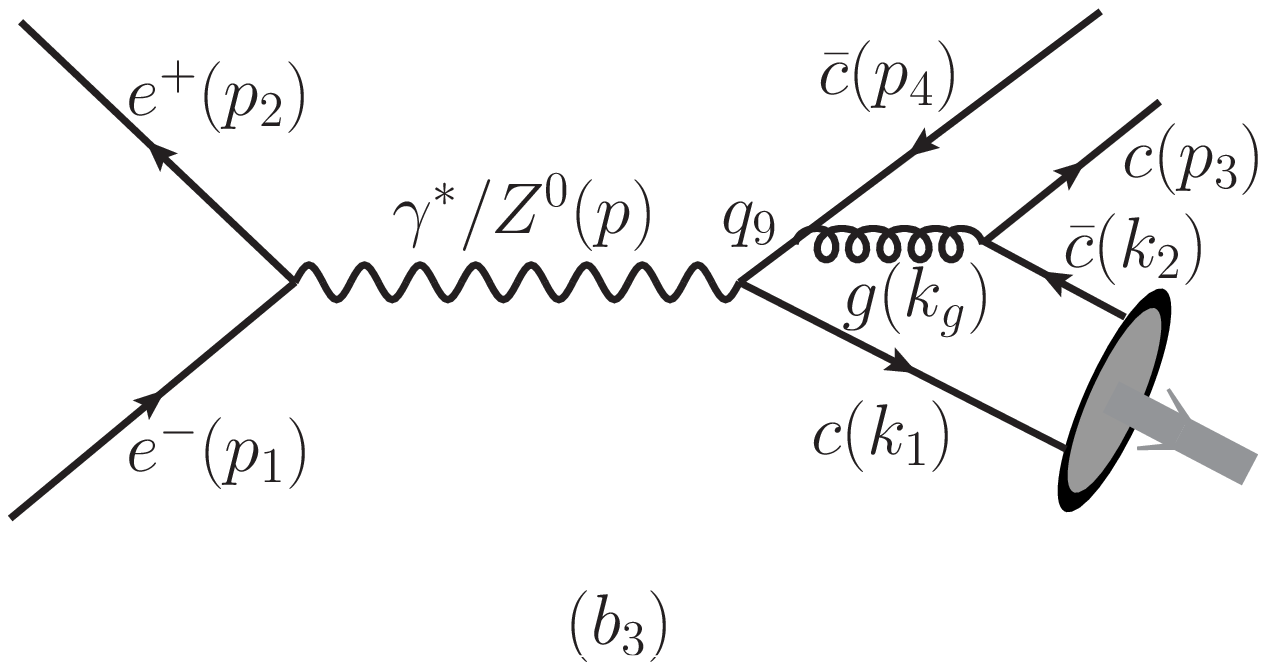}}
  \hspace{1in}
{
    \label{fig:subfig:10} 
    \includegraphics[width=0.35\textwidth]{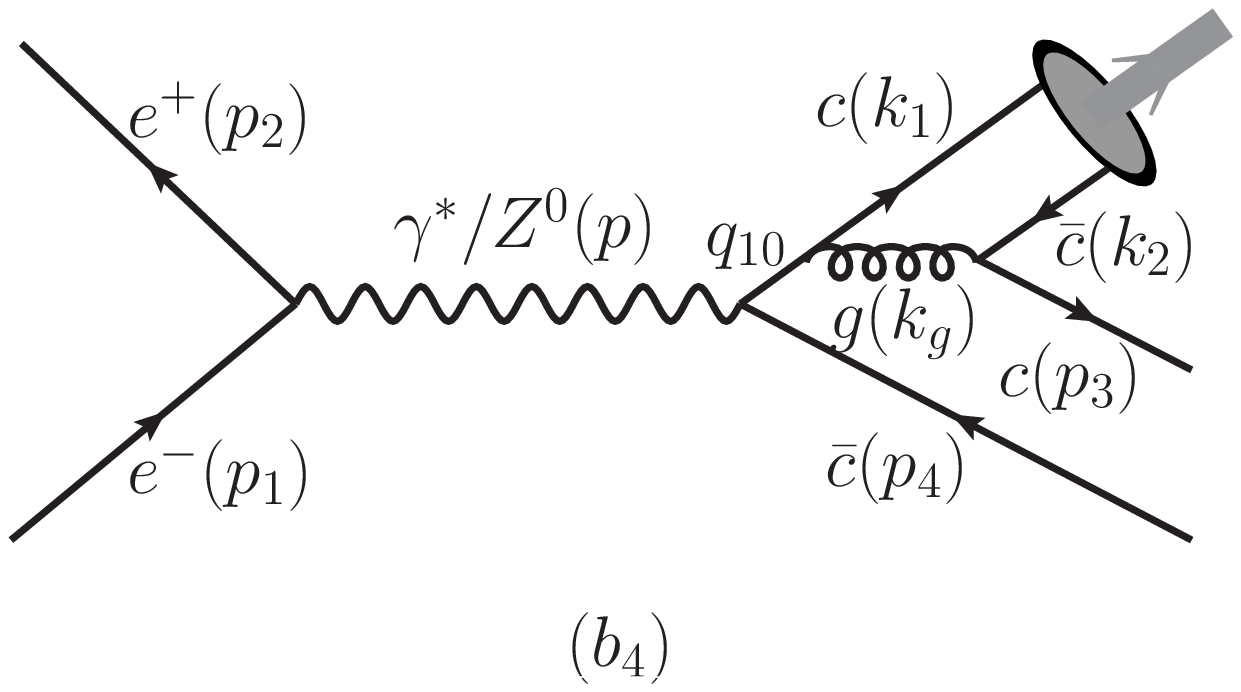}}
  \hspace{1in}
  \caption{Feynman diagrams for $P_c$ produces associated with a quark pair $q\bar q$ }
  \label{feynfignext} 
\end{figure}

In the following, we discuss the case that $P_c$ produces associated with a heavy quark pair $c\bar c$ ($b \bar b$), as shown in Fig. \ref{feynfignext}. The differential cross section for this kind of processes can be written as
\begin{eqnarray}
d\sigma&=&\frac{1}{2s}\frac{d^3k}{(2\pi)^3}\int\frac{d^3p_3}{(2\pi)^3 2E_3}\frac{d^3p_4}{(2\pi)^3 2E_4}(2\pi)^4 \delta^4(p_1+p_2-k-p_3-p_4)\nonumber\\
&&\times\int\frac{d^4k_1}{(2\pi^4)}\frac{d^4k_3}{(2\pi^4)} A_{ij}(k_1,k_2,p_3,p_4)(\gamma^0A^\dag(k_3,k_4,p_3,p_4)\gamma^0)_{kl}\nonumber\\
&&\times\int d^4x_1d^4x_3e^{-ik_1\cdot x_1+ik_3\cdot x_3}\langle0|\bar{Q}_k(0)Q_l(x_3)|P_c+X_N\rangle\langle P_c+X_N|\bar{Q}_i(x_1)Q_j(0)|0\rangle.
\end{eqnarray}
Similar to the above derivation, we obtain
\begin{eqnarray}
d\sigma&=&\frac{1}{2s}\frac{d^3k}{(2\pi)^3}\int\frac{d^3p_3}{(2\pi)^3 2E_3}\frac{d^3p_4}{(2\pi)^3 2E_4}\int\frac{d^4k_1}{(2\pi)^4} \frac{d^4k_3}{(2\pi)^4 }A_{ij}(k_1,k_2,p_3,p_4)(\gamma^0A^\dag(k_3,k_4,p_3,p_4)\gamma^0)_{kl}\nonumber\\
&&\times\int d^4x_1d^4x_2d^4x_3 e^{-ik_1\cdot x_1-ik_2\cdot x_2+ik_3\cdot x_3}\langle0|\bar{Q}_k(0)Q_l(x_3)a^\dag({\bf k})a({\bf k})\bar{Q}_i(x_1)Q_j(x_2)|0\rangle.
\end{eqnarray}
With the results in Eq.(\ref{eqv0}), the differential cross section for this kind of processes can be further expressed as
\begin{align}\label{NLOsigma}
  d\sigma &=\frac{1}{2s} \int \frac{d^3 k}{(2\pi)^3 v_0} \int \frac{d^3 p_3}{(2\pi)^3 2E_3}\frac{d^3 p_4}{(2\pi)^3 2E_4}(2\pi)^4 \delta^4(p_1+p_2-k-p_3-p_4)A_{ij}\Big[\gamma^0A^+\gamma^0\Big]_{kl}\nonumber\\
&\times\bigg\{-h_{1}(P_-\gamma_5P_+)_{ji}(P_+\gamma_5P_-)_{lk}+h_{3}\Big[(P_-\slashed vP_+)_{ji}(P_+\slashed vP_-)_{lk}-(P_-\gamma_{\sigma}P_+)_{ji}(P_+\gamma_{\sigma}P_-)_{lk}\Big] \bigg\}.
\end{align}

\begin{figure}[htb]
\centering
\scalebox{0.5}{\includegraphics{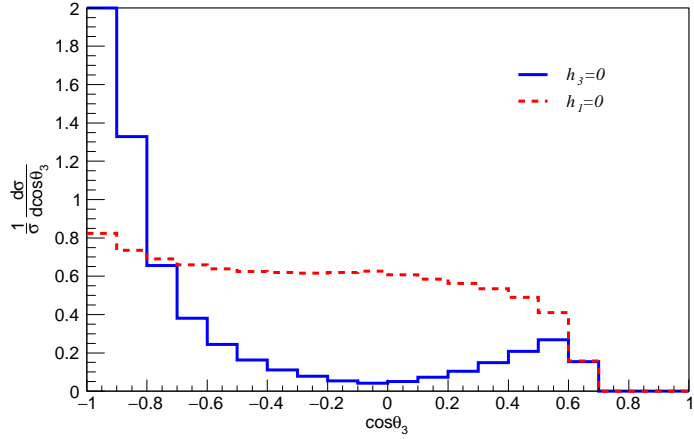}}
\caption{ Angular distribution $\frac{1}{\sigma} \frac{d\sigma}{d\cos\theta_3}$  of the $P_c$ production associated with a heavy quark pair at  $\sqrt{s}=91.2$ GeV. The solid line is for $h_3=0$, and the  dash one is for $h_1=0$. }
\label{cos3hist}
\end{figure}

The angular distribution $\frac{1}{\sigma} \frac{d\sigma}{d\cos\theta_3}$ at $Z^0$ pole is shown in Fig. \ref{cos3hist}, where $\theta_3$ is the angle between the momentum of the $c$ quark $(p_3)$ and that of $P_{Q}$. Because of mass effect, the cross section drops in the small angular range. From the figure, one can also note that for the case of spin-singlet ($h_3=0$), the cross section is suppressed when the angular $\theta_3$ is near $\frac{\pi}{2}$. In this case, the fragmentation processes as shown in the diagrams ($b_1$) and ($b_2$) of Fig. \ref{feynfignext} are prohibited because of the momentum conservation at the $g-q\bar{q}$ vertex. In the allowed cases shown by other Feynmann diagrams in Fig. \ref{feynfignext}, one must require one of the free charm quarks takes a relatively large transverse momentum with respect to the c-quark to compensate that of the $P_Q$.

Employing Eq. (\ref{NLOsigma}), we can get the total cross section of pentaquark production at B factory energy and at $Z^0$ pole. The numerical results of spin-singlet and spin-triplet contributions are given in Table \ref{Allrst}.

\begin{table}[htb]
\caption{Spin-singlet and spin-triplet contributions to the cross section of pentaquark production associated with a heavy quark pair $Q\bar Q$ at $e^+e^-$ colliders in units of fb. Here we take the values $h_1=h_3=0.0036$ GeV$^3$.}\label{Allrst}
\centering
\begin{tabular}{c|c|c|c|c}\hline
\multirow{2}{*}{Associate $Q\bar{Q}$}&\multicolumn{2}{c}{$\sqrt{s}=10.6$ GeV} & \multicolumn{2}{c}{$\sqrt{s}=91.2$ GeV}\\\cline{2-5}
&~~~Singlet~~~& ~~~Triplet~~~& ~~~Singlet~~~& ~~~Triplet~~~\\\hline
$c\bar c$  & 0.198 & 0.666& 0.0756 &  302.407   \\
$b\bar b$  & --  & --   & 0.0504 &  373.176   \\
\hline
\end{tabular}
\end{table}

Finally, we investigate the $P_c$  production in light quark ($u,d,s$) jet fragmentation.  It is reliable and straightforward at the PQCD level because of infrared safe. At lowest order,
the PQCD process is the gluon splitting into $c \bar c$ pair, and the Feynman diagram is shown in the diagrams ($b_1$) and ($b_2$) of Fig. \ref{feynfignext} with the free $Q \bar Q $ replaced by a light quark pair. The numerical results are presented in Table \ref{Allrst2}. One notices that the cross section for  this process at $Z^0$ pole is  comparable to the production  associated with a heavy quark pair and is larger than  the process $e^+e^-\to P_c+g$. In the latter process, because of momentum conservation, the hadron and gluon produced in the final state must recoil from each other, which makes the gluon very hard and it is suppressed.

\begin{table}[htb]
\caption{Spin-singlet and spin-triplet contributions to the cross section of pentaquark production from a light quark jet fragmentation in units of fb. Here we take the values $h_1=h_3=0.0036$ GeV$^3$.}\label{Allrst2}
\centering
\begin{tabular}{c|c|c|c|c}\hline
\multirow{2}{*}{Associate $q\bar{q}$}&\multicolumn{2}{c}{$\sqrt{s}=10.6$ GeV} & \multicolumn{2}{c}{$\sqrt{s}=91.2$ GeV}\\\cline{2-5}
&~~~Singlet~~~& ~~~Triplet~~~& ~~~Singlet~~~& ~~~Triplet~~~\\\hline
$u\bar u$  & 0.0 &  0.612  &  0.0  &   299.808   \\
$d\bar d$  & 0.0 &  0.155  &  0.0  &   399.636   \\
$s\bar s$  & 0.0 &  0.155  &  0.0  &   386.424  \\
\hline
\end{tabular}
\end{table}

To summarize, in this paper, we study the compact hidden-charm pentaquark production via the color-octet charm-anticharm pair fragmentation in $e^+e^-$ annihillation with clean backgrounds. The most straightforward application of our analysis is the  B factory at present and in the future. Based on our above calculations, at B factory  energies, the dominant production process is $e^+e^-\to P_c+g$. This means $P_c$ is dominantly produced in a two-jet like event. Belle collaboration now has collected an integrated luminosity about 1000 fb$^{-1}$ and the events number of $P_c$ production could be $10^5$. The future operation of B factory will accumulate even more events. So to set a jet algorithm trigger as suggested in \cite{Jin:2014nva} may help us to obtain a clear signal or upper limit for $P_c$ production in $e^+e^-$ annihillation. At high energies, e.g., in the future high luminosity Z-factory, since more rich partonic processes  available, as studied above, the observation is also possible. Once the direct production of the hidden charm pentaquark states is confirmed at $e^+e^-$ colliders, it will be very helpful to understand the quark model and the strong interactions.

\section*{Acknowledgements}
We greatly thank Profs. Yi Jin and  Zhong-Juan Yang for helpful discussions. This project is supported by National Natural Science Foundation of China under Grants No. 11635009 and 11325525.

\end{document}